\documentstyle[epsfig,12pt]{article}
\begin{document}
\def\be{\begin{equation}}
\def\ee{\end{equation}}
\def\beq{\begin{equation}}
\def\eeq{\end{equation}}
\def\eq{\beq\eeq}
\def\bea{\begin{eqnarray}}
\def\eea{\end{eqnarray}}
\newcommand{\nn}{\nonumber}
\newcommand{\lsim}{\raisebox{-0.13cm}{~\shortstack{$<$ \\[-0.07cm] $\sim$}}~}
\newcommand{\Lam}{\Lambda}
\newcommand{\eps}{\epsilon} 
\newcommand{\MS}{\overline{MS}} 
\newcommand{\qq}{\langle \bar q q \rangle} 
\newcommand{\CSB}{\raisebox{.8mm}{$\chi$}SB }
\thispagestyle{empty}
\begin{flushright}
PM/00--25 \\
hep-th/0107073
\end{flushright}
\begin{center}
{\Large\sc {\bf Borel convergence of the  variationally improved
mass expansion and dynamical symmetry breaking}}\\
\vspace*{10mm}
{\large J.-L. Kneur and D. Reynaud}\\
\vspace*{.5cm}

Physique Math\'ematique et Th\'eorique, UMR-5825-CNRS, \\
Universit\'e Montpellier II, F--34095 Montpellier Cedex 5, France.

\end{center}
\vspace{1.5cm}
\begin{abstract}
\setlength{\baselineskip}{16pt}
\noindent 
A modification of 
perturbation theory, known as delta-expansion 
(variationally improved perturbation), 
gave rigorously convergent series in some $D=1$ models (oscillator
energy levels) with factorially divergent ordinary perturbative expansions. 
In a generalization of variationally improved perturbation appropriate to 
renormalizable asymptotically free
theories, we show that the large
expansion orders of certain physical quantities are similarly
improved, and prove the Borel convergence of the corresponding series
for $m_v \lsim 0$, with $m_v$ the new (arbitrary) mass perturbation
parameter. We argue that non-ambiguous estimates of quantities 
relevant to dynamical (chiral) symmetry breaking
in QCD, are possible in this resummation framework.
\end{abstract}
\newpage
\setcounter{page}{1}
\section{Introduction}
A ``first principle" determination of 
the order parameters  
characterizing dynamical (e.g. chiral) symmetry breaking 
(\CSB) 
in asymptotically free theories (AFT) like QCD is traditionally
considered inaccessible (except perhaps from lattice calculations), 
due to three main obstacles:\\ 
\noindent $(i)$ order parameters, 
like the quark condensate $\qq^{1/3}$, are expected to 
be of ${\cal O}(\Lam_{QCD}) $, 
so that the coupling
at such scale is 
large: ordinary perturbative expansion is invalidated.\\
\noindent $(ii)$ At arbitrary perturbative order, $\qq$ and other 
\CSB order
parameters vanish anyhow in the massless limit: their chiral limits are
(perturbatively) trivial.\\
\noindent $(iii)$ A more subtle but equally important argument is
that, attempts to extract
genuine non-perturbative contributions to such quantities 
meet
inherent ambiguities, as indicated by the (infrared) renormalon
singularities of perturbative expansions\cite{David,renormalons}.
Conventional  wisdom thus treats 
$\qq$ and other non-perturbative condensates 
as {\em parameters} of a systematic
operator product expansion (OPE)\footnote{Unlike
the gluon condensate, the presence
of \CSB 
condensates like $\qq$ in OPE's 
is however not directly inferred by
infrared renormalons, 
these being screened by chiral
symmetry\cite{David}, cf. argument (ii) above. 
We will see that renormalons and argument (iii) 
are nevertheless relevant to \CSB quantities in our context.}, 
as best illustrated in the 
SVZ formalism\cite{svz}.\\
Yet in many field theory models, 
definite non-perturbative results may be obtained from an appropriately 
resummed (but different) expansion, 
like the 
$1/N$ expansion\cite{NJL,GN}.
There also exist powerful summation techniques,
like the Borel method\cite{Borel,renormalons} which, even for non
Borel-summable  expansions like
in QCD typically, gives nevertheless precious informations on the nature of
(power-like) non-perturbative contributions
to a given physical quantity.  
An alternative summation method, known as delta-expansion 
(DE) or ``variationally improved perturbation" 
(VIP)\cite{pms,delta}, is based
on a reorganization of the interaction Lagrangian to depend on arbitrary
adjustable parameters, to be fixed by some optimization prescription.
In various models DE-VIP exhibits (though often rather empirically) an 
improved convergence of perturbative expansion. Moreover
in some $D=1$ models, the anharmonic oscillator typically,
DE-VIP is equivalent\cite{deltaconv} to the ``order-dependent
mapping" (ODM) method\cite{odm}, and optimization is 
equivalent
to a rescaling of the adjustable oscillator frequency
(mass) with perturbative order, which can essentially suppress the
factorial asymptotic behaviour of ordinary perturbative coefficients. 
Such a procedure was proven rigorously to 
converge\cite{deltaconv,deltac} (for
an adequately rescaled mass) toward the exact
result, e.g. for
the oscillator energy levels\cite{ao} and related quantities.

Here we reconsider
a variant of DE-VIP adapted to higher
dimensional renormalizable theories, proposed some time
ago\cite{gn2}--\cite{qcd2}.  The basic idea is to perform a modification
of perturbative expansions in two stages: first exploiting
specific renormalization group (RG) properties, which transform the
ordinary expansion (in a coupling $g$) of certain physical
quantities, depending only on $g$ and on a mass $m$,
in the alternative form  
of ``mass power" expansions (MPE) in $(\hat
m/\Lambda)^{\alpha}$ [$\hat m$ is the renormalization  
scale-invariant mass, $\Lambda$ the
basic RG scale and $\alpha$ is given
by known 
RG coefficients]. This construction resums RG dependence to all
orders (at least in specific schemes), and most interestingly exhibits a
non-trivial massless (chiral) limit\cite{qcd1,qcd2} for DSB
(\CSB) order parameters, or for analogous quantities like the
``mass gap"  in $D=2$ models\cite{gn2}. 
However, such a result turns 
out to be well-defined 
only in the approximation of neglecting all the purely perturbative (non RG)
dependence. When arbitrary large orders  of the complete (non-log)
perturbative series are included, our naive mass gap result is plagued with
ambiguities originating mainly from renormalon singularities, cf. point (iii)
above, as we shall examine in more details here.  

However, in a second stage, an
appropriate version of the (order dependently rescaled) DE-VIP
can be performed on the complete MPE series in
$\hat m/\Lam$, essentially
replacing the true physical mass  by an
arbitrary adjustable mass parameter. In this note we mainly
investigate  the large order behaviour 
of the resulting ``variational" expansion 
in this mass parameter\footnote{See also 
ref. \cite{krqcd2001} for a preliminary discussion.}. 
We find that it produces a renormalization scheme (RS) dependent
factorial  {\em damping} of the original perturbative
coefficients at large orders. 
Yet, unlike the oscillator
case, the damping is generally not
sufficient to make the DE-VIP series readily convergent, 
in the standard perturbative regime,
when the
generically expected renormalon singularities are taken into account. 
But we show that the series can be {\em Borel} convergent,
if approaching the chiral limit with the 
arbitrary mass
parameter $Re[\hat m] \lsim 0$. These
results  apply formally a priori to any (asymptotically free) renormalizable
models.   Some concrete examples are the D=2 $O(N)$ Gross-Neveu (GN)
model\cite{GN} (where the mass gap is 
known exactly\cite{FNW}); 
or a $D=4$ gauged AFT with $n_f$ massless fermions like QCD, where the 
expected\cite{DCSB} $SU(n_f)_L \times SU(n_f)_R \to SU(n_f)_V$ breaking is
exhibited via non-perturbative order parameters.   
\section{Transmuted mass expansion and mass gap}   
\begin{figure}[htb]
\begin{center}
\vspace{-3cm}
\mbox{ 
\psfig{figure=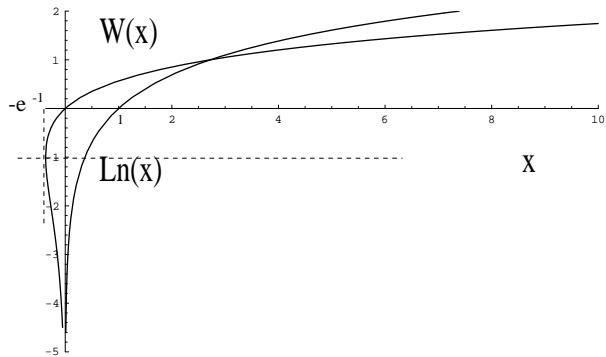,width=8cm}}
\end{center}
\vspace{-3.5cm}
\caption[]{\label{lambert} The Lambert function $W(x)$ compared to $\ln x$.}
\end{figure}

\begin{figure}[htb]
\begin{center}
\mbox{
\psfig{figure=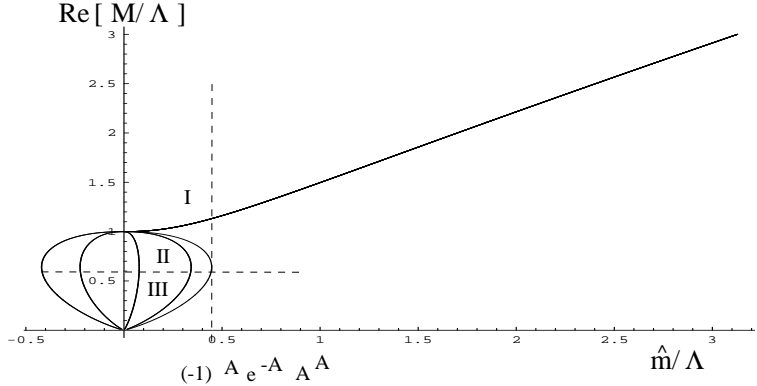,width=10cm}}
\vspace{-2.5cm}
\end{center}
\caption[]{\label{mongolfier} The different branches of 
$M_1$ in Eq.~(\ref{M1rewritten}), 
for $A=4/9$ (corresponding to first RG order QCD
with three active quark flavours).  }  
\end{figure}

In this and next section we essentially summarize some of the construction in
\cite{gn2}--\cite{qcd2}.
To illustrate simply the first stage, consider in a ``generic"
AFT
the first RG order evolution for the renormalized 
``current" mass:   
\beq 
M_1 = m(\mu) \; [1 +2b_0 g^2(\mu) \ln(M_1/\mu) 
]^{-\frac{\gamma_0}{2\,b_0}}\;,  
\label{MRG1} 
\eeq 
where $b_0$, $\gamma_0$ are one-loop RG coefficients,
with $b_0 >
0$ for an AFT [our normalization is $\beta(g) = -b_0 g^3
-b_1 g^5 -\cdots$,  $\gamma_m(g) = \gamma_0 g^2 +\gamma_1 g^4 
+\cdots$], and the   
{\em self-consistent} condition
$M_1 \equiv m(M_1) $
defines $M_1$. Now,
equivalently Eq.~(\ref{MRG1}) reads
\beq 
M_1 (\hat m) \:=\hat m \;[\ln (M_1/\Lam)]^{-A} \equiv \:\hat m \: F^{-A}  
\label{M1rewritten} 
\eeq  
with 
$\Lam = \mu \exp[{-1/2b_0 g^2(\mu)}]$  
the RG invariant scale, 
$\hat m \equiv  m(\mu) [2b_0 g^2(\mu)]^{-A} $ 
 the scale
invariant mass ($A \equiv \gamma_0/(2b_0)$), and 
in Eq.~(\ref{M1rewritten}) 
\be
F(\hat m/\Lam) \equiv
\ln (\hat m/\Lam) -A \ln F
\;= A\: W[A^{-1} 
(\hat m/\Lam)^{1/A} ]
\label{Fdef}
\ee 
where the Lambert\cite{Fproperties} function, $W[x] \equiv \ln x - \ln W$,
is plotted in Fig~\ref{lambert}. 
Eq.~(\ref{Fdef}) has the remarkable property: 
$F \simeq (\hat m/\Lam)^{1/A}$
for 
$\hat m \to 0$, in contrast with the ordinary Log (see Fig.~\ref{lambert}),
however asymptotic to $F(\hat m/\Lam)$ for $\hat m \gg \Lam$.
More precisely, on its principal branch (which is real-valued
for real arguments), $F$ has an alternative
series expansion:
\be
F(x) = \sum^\infty_{p=0} (\frac{-1}{A})^p 
\frac{(p+1)^p}{(p+1)!}\; x^{\frac{p+1}{A}}
\label{Fexp}
\ee 
of finite convergence radius $R_c = e^{-A} A^A$.
$M_1(\hat m)$ in Eq.~(\ref{M1rewritten})   
thus exhibits different branches
according to the values of the RG parameter 
$A$ (see Fig.~\ref{mongolfier}). Now, for most values of $A$, there is 
only one branch which for {\em real} $\hat m$ values, is real and
continuously matching
the asymptotic perturbative 
behaviour of $F$ at large $\hat m$: the one giving a 
non-zero ``mass gap" $M_1 = \Lam$ for $\hat m \to 0$ (region I in
Fig.~\ref{mongolfier}).  Algebraically, the mass is
obtained by expanding Eq.~(\ref{Fexp}) in
(\ref{M1rewritten}):   
\be
M_1(\hat m \to 0) = \hat m \;[(\hat m/\Lam)^{1/A}+\cdots]^{-A}\;
=\: \Lam \;(1+{\cal O}(\hat m/\Lam)^{1/A})\;,
\label{M1Lam}
\ee
which may be viewed as a generalization 
(for $m \neq 0$) of ``dimensional transmutation".
Note that Eq.~(\ref{M1Lam}) readily
reproduces, e.g., the GN $O(N)$ model mass gap in the large $N$,
$m\to 0$ limit (where $A\to 1$ for $N\to\infty$),
traditionally obtained in a different way~\cite{GN}.  
Eq.~(\ref{M1Lam}) is, however, not a proof of dynamical
\CSB: rather, {\em if} \CSB occurs, the  
fact that any mass is
proportional to $\Lam$ is consistently incorporated by the properties of 
$F(\hat m)$ for any $\hat m$, which provides an explicit bridge
between the ``non-perturbative" $\hat m \lsim \Lam$ regime, 
where $F$ has power
expansion (\ref{Fexp}),
and the short distance usual perturbative $\hat m \gg \Lam$ (Log) regime. 
A crucial point is the difference between the usual effective
coupling $g^2(p^2) \equiv 1/[b_0 \:\ln(p^2/\Lam^2)]$, having a Landau pole
at $p^2 = \Lam^2$, and $F^{-1}(\hat m)$ here, 
having its pole at
$\hat m =0$, governing the massless limit (\ref{M1Lam}) of the 
(pure RG) mass gap Eq.~(\ref{M1rewritten})\footnote{$W(x)$
appears in various branches of 
physics\cite{Fproperties}, in particular recently also 
in the QCD and RG context\cite{Grunberg}. 
Yet its connection with non-trivial 
chiral limit (\ref{M1Lam}) was unnoticed 
before \cite{gn2}--\cite{qcd2},
to the best of our knowledge.}. 
Accordingly along the continuous branch I, $M_1(\hat m)$ has
no singularity for $0 <\hat m <\infty$, as is clear
from Eq.~(\ref{M1Lam}) and Fig.~\ref{mongolfier}.     
%
\vspace{-2cm}
%
\vspace{2cm}
\section{Pole mass gap and other DSB quantities}
Eq.~(\ref{M1rewritten}) also defines a (lowest order) ``pole"
mass, being scale invariant to all orders 
(and gauge invariant as well, if gauge symmetry is relevant, as in QCD),
thanks to its continued fraction form in $M_1$.   
Yet the genuine pole mass is not given
simply by Eq. (\ref{MRG1}), as it includes  
non-log perturbative and RG contributions 
of arbitrary higher orders, 
though for most theories  one
only knowns at present the perturbative series up to the
second or third order coefficients, 
like e.g. in the GN model\cite{Gracey,gn2} or QCD\cite{Broadhurst}.\\ 
A generalization of (\ref{M1rewritten}), perturbatively consistent\cite{qcd2}
with the usual 
pole mass,
can be defined\footnote{Strictly, Eq.~(\ref{MRGn}) applies only
if $C \equiv b_1/(2b^2_0) \ge 0$. If $C <0$ (as in the $O(N)$ GN
model, corresponding to an infrared fixed-point at $g^2
=-b_0/b_1 >0$), an alternative appropriate RG summation 
can be defined\cite{gn2,KRnext}.}:  
\bea 
 M^P (\hat m) 
 = 2^{-C}\:\hat m  
F^{-A} [C + F]^{-B}   
\; \sum^\infty_{n=0} d_n\:(2b_0 F)^{-n}\;, 
\label{MRGn} 
\eea    
with
\beq 
F =\ln [\hat m/\Lam] -A  \ln F -(B-C) \ln [C +F], 
\label{F2def} 
\eeq 
\be
A =\frac{\gamma_1}{2 b_1},\:B =\frac{\gamma_0}{2 b_0}-A,\; C =
\frac{b_1}{2b^2_0}\;. 
\label{ABCdef}
\ee
$F(\hat m)$ in (\ref{MRGn}), (\ref{F2def}) resums the RG dependence in
$\ln[\hat m]$ at two-loop order exactly (or even to all orders in the scheme
$b_i =0, \gamma_i =0$ $\forall i\ge 2$).  
Most interestingly, similarly to Eq.~(\ref{Fexp}) $F$ also
has an ($A,B,C$ dependent) expansion  in $(\hat
m/\Lam)^{1/A}$ for sufficiently small $\hat m$, with $A$ now defined in 
Eq.~(\ref{ABCdef}).
The coefficients $d_n$ implicitly include
the non-log perturbative contributions from $n$-loop graphs
(generically dominant, as 
discussed below), plus eventually (subdominant) contributions from higher
RG orders.\\
A similar
construction can be performed for other physical quantities,
at least those depending only on $m$ and $g$.
Examples are  the perturbative expansion of the 
GN model vacuum energy\cite{gn2}, or in QCD 
the \CSB order parameters
$F_\pi/\Lam$ (the pion decay constant) 
and $\qq(\mu)/\Lam^3$\cite{qcd1,qcd2}.\\
Now in (\ref{MRGn}), 
there are crucial differences with the ``pure 
RG" mass gap, Eq.~(\ref{M1rewritten}):\\
--The pole mass (or other physical quantities similarly) 
is infrared finite, gauge~\cite{Tarrach}--, 
scale-- and scheme--invariant, 
but the relation between the pole mass and e.g. the running mass
is scheme dependent, which 
is manifested here by the RS-dependence in (\ref{MRGn})  
of: the perturbative coefficients $d_n$;
the RG coefficients $A$, $B$
in Eq.~(\ref{ABCdef});  
and of course $\Lam$. \\
--The 
dominant contributions $d_n$ in
(\ref{MRGn}) behave rather generically
as\cite{MPren,renormalons}:
\be
d_{n+1}
\raisebox{-0.4cm}{~\shortstack{ $\sim$ \\ $n\to\infty$}}
(2b_0)^n\: n!
\label{irren} 
\ee  
so that the series Eq.~(\ref{MRGn}) is badly divergent
for any $\hat m >0$, and not
even Borel summable: such a factorial growth of the
perturbative coefficients, with no sign alternation,
implies\cite{renormalons}
ambiguities of ${\cal
O}(\Lam)$, as we reexamine
within the present context in section 5. 
The $O(N)$ GN model mass
gap, at order $1/N$, also exhibits infrared 
renormalons similar\cite{KRnext}
to (\ref{irren}), if considering only its naive perturbative expansion. 
In QCD, insertions of the (resummed) gluon
propagator in the $F_\pi$ or $\qq$ perturbative expressions potentially 
give factorially growing asymptotic
coefficients: while usually considered irrelevant
in the $m\to 0$ limit (cf.
argument (ii) above), the factorial behaviour survives a priori in our
construction due to the non-trivial chiral limit\footnote{The form of those
``\CSB parameter renormalons" will be discussed
elsewhere\cite{KRnext,Dthese}.}.   
\section{Variationally improved mass expansion}  
We shall examine now how to possibly cure the
latter potential ambiguities of such a resummation of
 DSB quantities, by
combining the previous MPE series construction leading to 
e.g. Eq~(\ref{MRGn})
with a specific form of delta-expansion.
As mentioned in introduction, DE-VIP is essentially a
reorganization  of the interaction terms of the Lagrangian. 
More specifically here, we define a (linear) DE  as the substitution
\be 
m(\mu) \to (1-\delta)\;m_v; \; g^2(\mu) \to  \delta\:g^2(\mu) 
\label{substitution}
\ee  
within perturbative expressions at arbitrary order, where 
$m(\mu)$ is the renormalized Lagrangian mass 
(in e.g. $\overline{\small{MS}}$ scheme),
$\delta$ the new expansion parameter,
and $m_v$ an {\em arbitrary} adjustable mass. 
(\ref{substitution}) is equivalent to 
adding and subtracting to the massless Lagrangian a ``trial" 
mass term $m_v$ [$\delta$ interpolating between the 
free ($\delta=0$) and the
interacting  {\em massless} Lagrangian ($\delta=1$)], 
and is entirely
compatible with renormalization\cite{gn2} and  
gauge-invariance\cite{qcd2}.     
The procedure then usually\cite{delta} is to take the limit $\delta \to
1$ {\em after} performing a perturbative  
expansion of the relevant physical quantities 
to fixed order $\delta^k$, exhibiting a residual $m_v$ dependence, so that    
an optimization 
prescription, 
typically the ``principle of minimal sensitivity" (PMS)\cite{pms},
can be applied with respect to
$m_v$.  
However, we go here a step beyond this standard PMS usage 
by following more closely the logic that leads 
to rigorous convergence properties of the DE method for the oscillator. \\
In what follows we only investigate 
for simplicity the mass gap
Eq~(\ref{MRGn}), but our construction can easily
be generalized  to 
similar DSB quantities.
After applying substitution (\ref{substitution}),  
$M^{P}(\hat m, \delta) \equiv \sum_k a_k(\hat m) \delta^k$ can be most
conveniently directly resummed,
for $\delta\to 1$,
by contour integration\cite{gn2} around $\delta=0$, to arbitrary order $K$:
an appropriate change of variable
allowing to study the $m(\mu)\to 0$ (equivalently  
$\delta \to 1$) limit in Eq.~(\ref{substitution}) is: 
\be
\delta \equiv  1-v/K \;;\;\;\;    m_v =  K^\gamma \;\hat m_v\;.
\label{rescale}
\ee
Eq.~(\ref{rescale}) is simply a convenient
way of parameterizing how rapidly the Lagrangian mass $ m(\mu) \to 0$ limit
is reached (as controlled by $\gamma\le 1$) as function
of the (maximal) delta-expansion order $K$. Similarly to
refs.~\cite{deltaconv} the point is to adjust 
the rates at which 
$m(\mu) \to 0$ ($\delta\to 1$) and $K\to\infty$ are 
simultaneously reached, with no a priori need of invoking explicit
optimization principle.\\
\begin{figure}[t]
\begin{center}
\mbox{
\psfig{figure=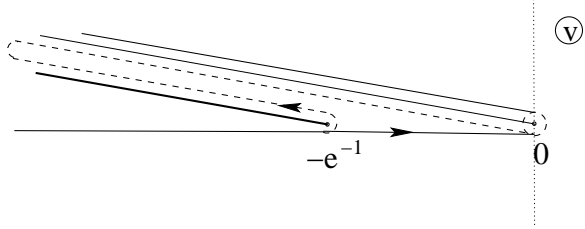,width=8cm}}
\end{center}
\caption[]{\label{contour} 
Singularities and equivalent integration contours in the
$v$ plane, for $A=1$.}  
\end{figure}
The final contour integral summation takes a simple
form, for $K\to\infty$:
\be
M^P/\Lam \sim \sum^N_{n=0}\frac{1}{2\pi i}\oint dv \;
e^{(v/m^")}\;F^{-A}[v]\:d_n\:(2b_0 F[v])^{-n}
\label{contour1}
\ee 
where $m^" \equiv \hat m_v/\Lam$, $N$ is maximal perturbative order,
and after deformation the contour encircles the semi-axis $Re[v]<0$ 
(see Fig.~\ref{contour}) and also 
for simplicity we fix from now the
scaling parameter in Eq.~(\ref{rescale}) 
to its maximal value ($\gamma =1$) still compatible with
massless limit (for $\hat m_v \to 0$). [The
general $\gamma$ scaling (\ref{rescale})
can be analyzed\cite{KRnext,Dthese} in a way more similar to the 
oscillator~\cite{deltaconv,deltac}, i.e. without the peculiar
contour $\delta$-summation Eq.~(\ref{contour1}),   
but largely complicates the algebraic analysis 
for renormalizable theories. 
In (\ref{contour1}) we also omit
some overall constant factors (due e.g. to $\Lam$ definition) 
irrelevant for
convergence properties,  and temporarily 
made a RS choice such that $B \equiv 0$
in (\ref{MRGn})--(\ref{ABCdef}), rendering 
certain algebraic expressions below more tractable,
without much loss of generality.]\\
Eq.~(\ref{contour1}) can be well
approximated analytically (at least for slightly restricted RS
choices, as indicated above and further below):
\be
M^P/\Lam \sim
1 +\frac{1}{2b_0}\;
\sum^N_{q=1} \left[\; \sum^{N-q}_{p=0} \frac{\Gamma[p+q] 
(p+q+A)(q+A)^{p-1}}{A^p\:\Gamma[1+p]\: \Gamma[1+q/A]}\; \right] (m^")^{-q/A} 
\label{Mpolesum}
\ee
where we assumed the leading
renormalon behaviour 
Eq.~(\ref{irren})\footnote{The original $n!$ 
coefficients in Eq.~(\ref{irren}) correspond
to $\Gamma[p+q]$ in (\ref{Mpolesum}). Higher order refinements on infrared
renormalon structure may easily be implemented: it essentially
replaces $(n-1)! \to  \Gamma[n +b_1/(2b^2_0)](1+r_{RS}/n+..)$ where 
 $r_{RS}$ depends on RS via $b_2$ etc~\cite{renormalons}, without affecting the
convergence properties discussed below.},   
 and we used essentially
Eq.~(\ref{Fexp}) together with   
$\oint dv e^v v^r = 2\pi i/\Gamma[-r]$ $\forall \:r $.\\
Now, some restrictions apply to  
(\ref{Mpolesum}): first, the sum over $p$ is bounded iff
\be
1/A \in \mathbf{N}^*
\label{Ainvint}
\ee
which we assume for simplicity from now. This is not
much restrictive, except that for arbitrary AFT it is generally 
not possible both that
$A$ satisfies (\ref{Ainvint}) {\em and}
$B=0$ in Eq.~(\ref{ABCdef}), as assumed in
(\ref{contour1}). But the more general scheme
$B\neq 0$ simply makes Eq.~(\ref{Mpolesum}) 
algebraically more involved, without affecting
the asymptotic behaviour and convergence properties discussed below.\\ 
Second, strictly (\ref{Mpolesum}) is
valid only asymptotically, for sufficiently
large $N$: due to the finite convergence radius 
of expansion (\ref{Fexp}), interchanging
the sum in (\ref{Fexp}) and integration in (\ref{contour1}) is not
rigorously justified. 
However, when (\ref{Ainvint}) holds, the formerly branch point $v=0$ is 
simply a pole, which allows to choose an equivalent 
contour of arbitrarily small radius around $v=0$, thus  
always inside the convergence radius of
(\ref{Fexp}) (see the dashed small circle
contour in Fig.~\ref{contour}). So, 
only the simple pole terms $v^{-1}$ contribute to 
Eq.~(\ref{contour1}), which finally sum up to
(\ref{Mpolesum}).  
The extra contribution (around the cut at $v =
-e^{-1}$, e.g. for $A=1$)
gives the difference between the ``exact"
integral (\ref{contour1}) and expansion (\ref{Mpolesum}), 
and can be evaluated numerically. These 
contributions are easily
shown for $A=1$  to contribute as
${\cal O}( e^{-(e\:m^")^{-1}}) h[N]$ 
relative to (\ref{Mpolesum}),
where $h[N]$ rapidly 
decreases for $N\to\infty$. Thus for 
large enough $N$ (and/or small $m^"$) 
those contributions are unessential for the convergence
properties discussed below.\\
The announced factorial damping of 
coefficients,
as compared
to the original perturbative expansion, is explicit in 
Eq.~(\ref{Mpolesum}). Yet, closer examination 
indicates that the damping is insufficient to make this series
for $N \to \infty$ readily convergent. Before considering the asymptotic
behaviour of the full series
(\ref{Mpolesum}), it is instructive to
examine the $p=0$ terms,  
behaving as:  
\be
\sim \sum^N_q \Gamma[q]/\Gamma[1+q/A]\;(m^")^{-q/A}\;.
\label{naive}
\ee
The denominators in (\ref{naive}) overcompensate
the numerator factorials iff 
\be
 0 < A \le 1 \;,
\label{Acons}
\ee
where for $D>1$ AFT, $A$ is RS dependent, as discussed in section 3.
Thus, if our series would only consist of terms of the form 
Eq.~(\ref{naive}),  
the solution would be simply
to perform appropriate scheme changes $A \to A^{'}$
in (\ref{MRGn}), (\ref{Mpolesum}) etc, so that a damping of coefficients 
larger than (or equal to) the factorial growth would make the series
convergent. [For such RS changes in $A$
one should consistently derive the corresponding change in e.g. the first few
perturbative coefficients $d_1$, etc, and in $\Lam$, but this one-parameter RS
change does not reintroduce any factorial behaviour in $d_n$ at large
orders. Moreover, if (\ref{Acons}) holds, any generic
infrared (or ultraviolet) renormalon
behaviour, of the form\cite{renormalons} 
$\sim r^n \,n!$ with $r$ arbitrary, is damped
similarly.]\\
Unfortunately, the large $N$ behaviour 
of (\ref{Mpolesum}) differs from the simple ``oscillator form" 
(\ref{naive}), due to 
the $p \ge 1$ terms in expansion (\ref{Fexp}) 
reminiscent of RG properties.
For any low $p \ll N$, renormalon
factorials are still overcompensated if $A \le 1$,
but the $\Gamma[1+q/A]$ damping decreases in strength
as $p$ increases, giving increasing contributions to the sum over $p$. 
All in all, the leading contributions 
to the coefficients of (\ref{Mpolesum}) happen at intermediate values of
$p$.
Nevertheless, the idea of damping factorials from appropriate
RS choice does survive, when the series Eq.~(\ref{Mpolesum}) 
is Borel transformed, as examined in next section. 
\section{Borel convergence of DE-VIP}
A Borel integral
slightly adapted to our case reads: 
\be
BI(\hat m) \equiv \tilde M^P(\hat m)   =  2^{-C}
\hat m\:F^{1-A}(C+F)^{-B}\int^\infty_0 \! dt e^{-F\:t}\;
[1 +(2b_0\,F)^{-1}\sum^\infty_{n=0} t^n \:] 
\label{BI}
\ee
which
would be (asymptotically) equal to (\ref{MRGn}) by formal
expansion\footnote{We define the Borel
transform (integrand of (\ref{BI}))
by dividing series coefficients by $(n-1)!$ for convenience. 
Also, the summed RG-dependence  $\hat m F^{-A}(C+F)^{-B}$,
having no factorial
behaviour, is
factored out of the Borel transform.} 
(upon assuming Eq.~(\ref{irren})), 
would the pole at  $t_0=1$ not make the integral (\ref{BI})
ill-defined. One should make a choice in deforming the contour e.g. above (or
below)  the pole,
which results in an ambiguity, easily calculated 
to be ${\cal
O}(e^{-F})$. Since $F \sim \ln [\hat
m/\Lam]$ for $\hat m \gg \Lam$, an ${\cal O}(\Lam/m)$ 
ambiguity\cite{renormalons} for the ``short distance" ($M, \hat m \gg
\Lam$) pole mass is recovered. But in our construction Eq.~(\ref{Fexp}) allows 
to trace the behaviour of $F$ all the way down to $\hat m\to 0$, where 
$F\to 0$: there
the ambiguity becomes ${\cal O}(1)$, and the naive RG--summed mass gap
(\ref{M1Lam}), which is $ {\cal O}(\Lam)$,
gets an ambiguity of same order, as announced.\\
Now for any given choice of contour avoiding the pole
(or cut\cite{renormalons} at higher RG order) in the Borel plane $t$,
let us apply the DE-VIP as defined in section 4,
introducing the $\delta$--expansion and contour resummation as
in (\ref{contour1}), 
this time on the Borel integral Eq~(\ref{BI}).
Interchanging the contour and Borel integrals, one can find
after some algebra the asymptotic behaviour for 
$N\to \infty$:
\be
\displaystyle
\tilde M^P_{var}(m^") \sim \Lam [ 1+  
\int^\infty_0 \!\frac{dt}{2b_0}\,\,
\sum^\infty_q 
\frac{(t^A e^t/m")^{q/A}}{\Gamma[1+q/A]}\;]
\label{BIvarbis}
\ee
where  we neglected for simplicity here the two-loop RG 
dependent $(C+F)^{-B}$ term in (\ref{BI}), as it does not affect asymptotic
behaviour.
It thus appears that the asymptotic behaviour of the Borel integrand 
in (\ref{BIvarbis}) is that of an entire series (at least for $A>0$), i.e. with
no poles for $0< t < \infty$. More precisely, the pole at $t_0=1$ in the 
original (standard) Borel integrand has been pushed to $t_0 
\to +\infty$ due to the factorial damping, so that the Borel integral is 
no longer ambiguous. However, integral 
(\ref{BIvarbis}) is not convergent, at least for $Re[m"]>0$, so that the 
series is not Borel summable for standard (perturbative) $m"$
values.\\  
But conversely, the integral in Eq.~(\ref{BIvarbis}) can converge,
for $Re[m"]<0$. This is the case at least for $A=1$, which can always be
chosen by an appropriate and simple RS change, as previously explained. 
Now, since $m"\equiv m_v/\Lam$ is an arbitrary parameter (and physical
quantities anyway only depend on $m^2$ in relativistic field theories), 
it should be legitimate to reach the chiral limit $m"\to 0$, of main interest
here, within the Borel-convergent half-plane $Re[m"]<0$. 
For $A\ne 1$, one may also
choose the arbitrary parameter $m"$ with
$Re[(m")^{1/A}] <0$ such that (\ref{BIvarbis})
converges, though this appears not possible for any arbitrary $A$ values. This
is however only an artifact of our simplest choice of the $\delta$-expansion
summation defining the DE-VIP series and leading to (\ref{BIvarbis}): 
for instance, an appropriate ($A$-dependent) generalization of
Eqs.~(\ref{substitution})-(\ref{rescale}) defining the DE-VIP expansion,
directly leads to a Borel convergent series independently of $A$
values\cite{KRnext}. Moreover, as already mentioned the function $F(\hat m)$
in (\ref{Fdef}) is well-defined (analytic) for any $A$ values
in a circle of radius $e^{-A} A^A$ around zero (and for $A=1$ the only
singularity is at $F=-1$ i.e. $\hat m/\Lam = -e^{-1}$, cf. Fig.~\ref{lambert}).
The higher RG order $F$ in  Eq.~(\ref{F2def}) has similar properties, 
with finite (but RS dependent\cite{qcd2}) convergence radius around
zero. Thus, one can choose $Re[m"] \lsim 0$ and/or equivalently 
$Re[F(m")] <0$, while
the mass gap $Re[M(F)] \equiv Re[M(\hat m)]$ always remains positive, see
Fig.~\ref{mongolfier}. \\
Actually, one can see directly the Borel
summability of Eq.~(\ref{BI}), independently of $A$, without
need of considering the DE-VIP expansion: if $F <0$,  $F \equiv -|F|$
simply produces the adequate sign alternation in the factorially
growing coefficients.  More precisely,  a straightforward
calculation of Eq.~(\ref{BI}) for $Re[F]<0$ (again neglecting the
two-loop RG dependence $C$, irrelevant to asymptotic properties), gives: 
\be
\tilde M^P/\Lam  \sim  
e^{-|F|}+\frac{1}{2b_0}\: Ei(-|F|)
\label{directBS}
\ee
where the exponential integral function $Ei(-x)$ has 
well-defined (sign alternated) asymptotic expansion for $x>0$. \\
Note however that the DE-VIP expansion, leading 
to Eqs.~(\ref{BIvarbis}), appears to improve further the 
asymptotic behaviour, at least for $A=1$, as compared to (\ref{directBS}), due
to the extra factorial damping. 
For $A=1$ and $Re[m"] \lsim 0$  
Eq.~(\ref{BIvarbis}) becomes after
integration   
\be
\tilde M^P_{var}(m^") \sim
const.\;\Lam\; (1+f(|m"|)\;)
\label{MPBsum}
\ee
where $f(|m"|)\to 0$ exponentially fast. The (here unspecified) overall
constant in (\ref{MPBsum}) originates essentially from RG dependence, involving
non-trivial factors such as $2^{-C}$ etc at second RG order, cf.
Eq.(\ref{MRGn}), (\ref{BI}).\\ 
We have thus obtained Borel convergence for a certain range of the
arbitrary mass, strictly only for $Re[m"]<0$, but in which in addition
the purely perturbative contributions are small and even vanishing in 
the chiral $|m"|
\to 0$ limit.  This does not mean, though, that our final DE-VIP result is
completelly independent of the perturbative information, since the above
series are only asymptotic to the exact series. Rather, it suggests that
the ``non-perturbative" result in the chiral limit may
be essentially determined by pure RG properties, plus eventually the very
first few perturbative terms, but not influenced by details of the large
perturbative orders. Note indeed
that, only from the properties of $F$ around $F \lsim 0$
(thus independently of the DE-VIP construction) the Borel sum
in (\ref{directBS}) 
reproduces at least qualitatively
the asymptotic behaviour of the {\em exact} 
$1/N$ result in the $O(N)$ GN model\footnote{The Borel summability of the exact
$1/N$ $O(N)$ GN model mass gap, independently of the present construction, is
analysed in details in \cite{KRcancel}.}: in this model the $2b_0$ in
(\ref{directBS}) is more precisely replaced by $N-2$, and the exact
result\cite{FNW} has an asymptotic expansion\cite{KRcancel} similar to
(\ref{directBS}), except for a finite term $\gamma_E$, which 
not suprisingly cannot be guessed
by our simple Borel summation of the (leading) renormalon behaviour
in Eq.~(\ref{BI}). 
\section{Discussion}
Though renormalon ambiguities 
are perturbative artifacts expected to disappear 
(or more precisely to cancel out with OPE contributions) 
in truly non-perturbative
calculations\cite{David,renormalons,Benbraki,KRcancel}, such explicit
cancellations  are generally inaccessible
for theories like QCD.  
Rather, the peculiar damping mechanism
of factorial divergences exhibited here is intuitively  
due to the fact that our reorganization of perturbative expansions
makes those much more similar to the oscillator energy levels expansion,
exhibiting a dependence on $\hat m_v/\Lam$, Eq.~(\ref{Fexp}), which is 
{\em power-like}
(rather than log-like) for sufficiently small $\hat m_v$. Moreover, 
the adjustable parameter $\hat m_v/\Lam$ may be order-dependently
rescaled, or can take arbitrary values, in particular $Re[\hat m_v/\Lam]<0$
producing sign alternation of factorial coefficients.
The DE-VIP expansion appears in that way to "bypass" the need for
explicit (and generally complicated) cancellation between perturbative and
non-perturbative contributions, at least for certain physical quantities like
the mass gap.  
Note also that the linear DE-VIP taking the form (\ref{contour1}), and 
(\ref{BIvarbis}) 
when combined with the Borel method, is only one among other
possible similar resummations. In particular,
we emphasize that the obtained convergence 
properties  for $Re[F] <0$ ($Re[\hat m_v/\Lam]<0$)
do not depend on the detailed properties of the ``delta-expansion" contour
integrals here considered, e.g. Eq.~(\ref{contour1}) [which lead
however to a rather simple and tractable
expressions in the massless limit and for Borel transforms
Eqs.~(\ref{BIvarbis})--(\ref{MPBsum})]:  
more generally  applying the 
$\delta$-expansion idea in slightly different forms may
replace (\ref{Mpolesum}) and subsequent results 
with eventually different series~\cite{KRnext}, but with 
similar asymptotic and 
(Borel) convergence properties.\\
In summary,  our construction exhibits
an explicit counter-example to 
conventional wisdom arguments
(i)--(ii), and to some extent (iii), mentioned in introduction. In the present
paper we have only analyzed the formal Borel
convergence properties, a priori applicable 
to any AFT, relying essentially
on the properties of $F(\hat m)$ in Eqs.~(\ref{Fdef}),(\ref{F2def}),
interpolating smoothly from the ordinary perturbative coupling to
the infrared mass power expansion.
These convergence properties can 
be viewed as the generalization to $D>1$ 
renormalizable theories of the ordinary
convergence properties of the DE-VIP for
the oscillator\cite{deltaconv,deltac}.   
Next we argue 
that such a summation recipe can
provide a well-defined basis 
to estimate more precisely some of the
\CSB order parameters in QCD or other models,
and a more detailed study 
with concrete numerical applications to 
the GN model and QCD will
be explored in \cite{KRnext}.
Though one may eventually raise that in QCD-like theories, 
other contributions  to the \CSB order parameters 
of ``truly non-perturbative" origin 
(i.e. unreachable by any resummation mean,
and/or related e.g. to instanton phenomena typically) may be expected, the
resummation contributions here considered should be a useful
piece of information.\\

{\bf Acknowledgements}\\
We thank Andr\'e Neveu for valuable discussions related to this work,
and Martin Beneke and Georges Grunberg for useful
previous discussions on the general properties of renormalons.
\end{document}